# HPC with Enhanced User Separation

Andrew Prout, Albert Reuther, Michael Houle, Michael Jones, Peter Michaleas,
LaToya Anderson, William Arcand, Bill Bergeron, David Bestor, Alex Bonn, Daniel Burrill,
Chansup Byun, Vijay Gadepally, Matthew Hubbell, Hayden Jananthan, Piotr Luszczek,
Lauren Milechin, Guillermo Morales, Julie Mullen, Antonio Rosa, Charles Yee, and Jeremy Kepner

*Massachusetts Institute of Technology*
{aprout,reuther,michael.houle,michael.jones,pmichaleas,
latoya.anderson,warcand,bbergeron,david.bestor,alexander.bonn,daniel.burrill,
cbyun,vijayg,mhubbell,hayden.jananthan,piotr.luszczek}@ll.mit.edu,
milechin@mit.edu, {guillermo.morales,jsm,antonio.rosa,yee,kepner}@ll.mit.edu

*Abstract*—HPC systems used for research run a wide variety of software and workflows. This software is often written or modified by users to meet the needs of their research projects, and rarely is built with security in mind. In this paper we explore several of the key techniques that MIT Lincoln Laboratory Supercomputing Center has deployed on its systems to manage the security implications of these workflows by providing enforced separation for processes, filesystem access, network traffic, and accelerators to make every user feel like they are running on a personal HPC.

## I. INTRODUCTION

Throughout the history of high performance computing (HPC) security has needed to be integrated very carefully with the overall system design to ensure that the performance goals of the system can still be met. While there has been much focus recently on application and system development teams including security considerations early in the design of programs, HPC system operators never had the luxury of moving away from that model in the first place. Some of the early security/performance trade-offs considered acceptable for HPC systems are no longer acceptable as the userbase of HPC has widened, multi-tenancy becomes more common, and insider threat concerns have become more acute. For HPC systems to achieve their performance goals, the tolerances on a wide variety of system actions are much tighter than on desktop or enterprise computing, and deviations can have cascading effects [1]. What might be a trivial or acceptable impact on a desktop computer can be a catastrophic impact to performance on an HPC system. A few milliseconds longer to load a web-page is trivial, while a few milliseconds longer for a remote dynamic memory access (RDMA) transfer can significantly degrade a message passing interface (MPI) job. Solutions that work on systems with a small number of CPU cores can bottleneck with manycore systems that are common in HPC. It is true that some security features/patches do degrade performance, but how and where they do so are important to HPC. For example, the patches that mitigated the Spectre and Meltdown hardware vulnerabilities impacted performance between 15-40% [2]. Some industries were unwilling to pay this penalty and operating systems switches were added to run in "insecure" mode [3]. Additionally, HPC systems have at times shunned the concept of defense in depth as the trade-offs of multiple security controls for the same activity, with the associated multiple performance impacts, was considered unjustifiable. This has left systems brittle to novel vulnerabilities [4]. There are obviously more examples like that.

However, there are plenty of other security measures that do not have performance impact. These measures may have other impacts on the users and operations of the HPC system. We have been exploring these trade-offs for more than two decades on the HPC systems we design, build and operate at the MIT Lincoln Laboratory Supercomputing Center (LLSC). Since its early days, the MIT LLSC has focused on developing a unique, interactive, on-demand HPC environment to support diverse science and engineering applications. This system architecture has evolved into the MIT SuperCloud [5]. The MIT SuperCloud has spurred the development of a number of cross-ecosystem innovations in interactive HPC [6], [7], HPC education [8], high performance databases [9], [10], database management [11], data protection [12], database federation [13], [14], data analytics [15], dynamic virtual machines [16], [17], distributed web applications [18], system monitoring [19], and the economics of operating HPCs [20]. This capability has grown in many dimensions. MIT SuperCloud not only continues to support parallel MATLAB and Octave jobs, but also jobs in Python, Julia, R, TensorFlow, and PyTorch along with parallel C, C++, Fortran, and Java applications with various flavors of message passing interface (MPI), scientific, numeric, and data analysis libraries. It is in this highly flexible HPC prototyping system/environment that we have been exploring various security measures and features, all while working hard to not affect application performance.

To start out, we need to ask ourselves the question: Why should we concern ourselves with securing HPC systems? HPC systems are probably not targets, right? And the current security is probably good enough for the low priority they are for being exploited, right? The answer to those questions is a resounding "No!" First, the importance of the security of our information systems became abundantly clear with the Edward Snowden leaks in the United States, which prompted amplified efforts worldwide to more effectively secure all of our information systems. But more succinctly regarding HPC systems, there have been numerous compromises and breaches of HPC systems. The most egregious of these breaches occurred across numerous European HPC centers in May 2020; HPC centers in Germany, Switzerland, United Kingdom, and possibly Spain were taken offline for hours and perhaps days to recover and clean up from the breach [21]. These breaches seriously affected COVID-19 simulation and data analysis efforts. But the potential for loss is not only delays of research projects, though these losses can be dramatic in and of themselves. There is also potential risk for economic loss when an HPC system has to be rebuilt and no jobs are being run, risk of not getting new grants and projects, and risk of no longer being able to trust the integrity of data and computations on the HPC system. Finally, it is likely that HPC system security requirements will be written into future government funded project requirements in many countries across the world.

## II. How is HPC Different?

HPC systems are quite different from conventional enterprise installations in some significant ways [22]. First, enterprise systems and HPC systems both have a huge number of files contained in their central file systems with a large number of users accessing some subset of the files consistently every day. However, the files on the enterprise storage systems change slowly, at human editing pace, and the changes are generally localized, multiple changes to the same file in quick succession. On HPC systems, many files are changing simultaneously and quickly, mostly from executing parallel jobs, and the changes are occurring in many places in the file system corresponding to the present working directory of currently executing parallel jobs. Second, the users of enterprise systems are usually from the same organization, who are all under the same rules and requirements and are aligned toward common organizational goals. For HPC systems, it is not uncommon for the users to be from multiple organizations using the system to collaborate on common projects; however, they differ on their organizational rules and goals. Finally, enterprise systems host a homogeneous set of applications: users run a limited number of sanctioned applications on their own computers such as office applications, browsers, email clients, etc. Also, business applications and databases are usually centralized in the enterprise.

The greatest difference between enterprise users and HPC users is that every HPC user is a software developer, but software development is not primary domain of expertise for most users! In fact, very few HPC users have workflows that don't require them to write code. This can present in many different ways: writing algorithms in Python/Julia/Matlab/Octave, setting up processing pipelines, performing analysis, creating multi-workflow orchestration via shell scripts, developing a complex distributed simulation using C and MPI, etc. At the very least, users create scenario files or parameter files for executing simulations or data analyses, which affect how the application executes. Some of the code is early prototype "version 0" code, which is going to have bugs and is not going to have any security built in (yet). Even venerable HPC libraries have little security built-in. For instance, MPI frameworks do not encrypt data or authenticate peer ranks, and efforts to extend them with security have seen little adoption [23], [24]. And users are often required to run software from large open source frameworks and/or proprietary closed-source programs, neither of which are typically designed with an HPC environment in mind. Both have unique challenges making them difficult or impossible to modify to better suit this environment. Further to the point, for researchers, software is not the end product. In many cases, the primary goal of running the simulation, data analysis, AI model, etc. is to generate data that will appear in a plot in a presentation or as text in a report. Obviously, there are security concerns running these codes including:

- It may be interacting with sensitive data;
- It is usually distributed across multiple compute nodes and computing over the HPC network; and
- It is executing on a shared use system.

All of these security concerns require that mitigations are put into place to reduce or eliminate security risks.

## III. Path Forward

We have established that HPC systems are different from conventional enterprise computing systems, and that the greatest difference is that every HPC user is a software developer. And that begs the question: How do we manage this risk? There are two options.

Option #1 is to make the code better, by focusing on improving the most commonly-used software and development libraries and frameworks, thereby providing easy to use security primitives and training users to be better, security-focused programmers. There are a number of challenges with this approach:

- It does not solve the issue of where to run "version 0" of code.
- It does not address large open source frameworks or closed-source commercial software.
- There is a daunting variety of software that is run on HPC systems, and we cannot fork every software project to create a "multi-tenant HPC" version.
- It still requires users to prioritize writing secure code and use any primitives that are provided.

- User turnover can be overwhelming. In particular, for university systems, a new class of students arrive every fall for academic HPC systems.

These challenges strongly suggest that this option is unattainable. But there is a second option.

Option #2 involves making the HPC system itself more secure. As we remarked before, every HPC user is a software developer, and they need a coding sandbox, a safe testbed for the initial development of new code and for when the code still has bugs and needs to be debugged (including security-relevant bugs). It would enable fast exploration of capability because not all coding efforts will turn into successful projects — some are intentional one-offs. The security benefits would extend even to much more mature code, because even software designed with HPC in mind rarely fully considers security. Can the HPC system be configured in such a way that each user has their own development sandbox? In the next section, we detail a set of configurations and technologies that provide such enhanced user separation capability to our userbase.

## IV. Implementation

Implementing enhanced user separation on a HPC[1] capability involves enhancing the separation between users and isolating them from each other so they cannot observe or interact with each other. The solution encompasses a number of configurations and measures that span the HPC system because there are many places within an HPC system where users can interact on the shared system or at least can observe the activities of other users. Throughout this Section, we assume that the operating system of the HPC system is a variant of Linux, but these configurations and measures can be applied to other Unix variants and possibly even to Windows-based clusters. The next subsections systematically step through each of the areas of an HPC in which user interactions can occur along with the security configurations and measures that are applied in that area. These areas are: processes; scheduler; filesystems (local/shared); networks; web applications; accelerators (GPUs, etc.); and containers.

### A. Processes

Having a shared use system means, by definition, that multiple users are simultaneously occupying and consuming resources of the system. By default, users on a shared use HPC system are allowed to view other users' processes, resource occupation, and utilization. But it doesn't have to be this way, and for many reasons, security and otherwise, it can be very beneficial to configure the system in such a way that users cannot see each others' processes, resource occupation, and utilization. We accomplish this by restricting visible Linux process information by setting `hidepid=2` on the `/proc/` mount. This isolates and hides processes and command line entries belonging to other users or system daemons, and it solves an entire class of information leakage issues. We benefited from this when SLURM CVE-2020-27746 (https://nvd.nist.gov/vuln/detail/CVE-2020-27746) was announced, as this configuration effectively mitigated the vulnerability in advance on our systems — the nirvana situation of security defense in depth. In addition to the security benefits, this also provides a better user experience because users only see the things they should care about, without the need to scroll or grep through a large amount of things they should never need to see.

[1]Enhanced user separation on a HPC is related to or called by other names depending on who you are talking to: multi-tenant HPC, multi-program, Protection Level 2 (PL2), or need-to-know separation.

On a related note, we have found that HPC support personnel who are not full administrators of the system (e.g., HPC research facilitators and HPC solutions architects) often need some escalated privileges. For instance, it is helpful for them to be able to view overall system load and attribute hotspots to specific users to help troubleshoot an execution script or a failed job execution. We have implemented a tool called `seepid`, which allows a whitelisted set of HPC support personnel to add a supplemental group to their logon session that is configured to be exempt from the `hidepid` restrictions via the `gid` flag on the `/proc/` mount.

### B. Scheduler

Most schedulers by default allow users to see other users queued and executing jobs. They may even be able to get job reports of any and all other users on the system with the submission of a single scheduler command. However, schedulers also have configurations with which such activities can be curtailed. In Slurm, it is also possible to isolate users from each other. The `PrivateData` configuration is used to restrict globally visible scheduler information, thereby hiding other users' jobs, usage, scheduling, information, accounting information, etc., which are accessible through the scheduler. This mitigates many of the same information leakage concerns as processes, since many job properties could contain private information including username, jobname, command, working directory path, etc.

We have found it worthwhile to go a step further. Several options are available in regard to whether individual compute nodes are shared simultaneously among users. By default, schedulers are configured to enable the jobs of multiple users to be executed on each compute node. This is one policy tactic that can enable greater job throughput and utilization, but it also has some trade-offs. For instance, if a node fails because one of the tasks executing on it tries to use more memory than is available on the node, all of the jobs running on that same node will fail. Schedulers do provide an option to request an exclusive run time environment for a job during the job submission; with exclusive job execution, only tasks from that given job are allowed to execute on the nodes on which that job has been dispatched to execute. This works for some situations, but it results in poor utilization if a user is executing many bulk synchronous parallel jobs like parameter sweeps and Monte Carlo simulations.

Recently, LLSC made a significant change to their scheduling policy to address these issues. LLSC has moved to a user-based whole-node scheduling policy where whole compute nodes are allocated to each user [25], [26]. In other words, once a user's job is dispatched to a compute node and there are unscheduled resources still available on that node, only other jobs from that same user can be scheduled on that node; jobs owned by other users cannot be scheduled on that node. This guarantees that, at any point in time, each compute node is only executing one or more jobs/tasks owned by one and only one user.

With the implementation of whole-node scheduling, one might remark that the process hiding detailed in the previous Subsection would be unnecessary. However, we would disagree. It is still best to hide OS processes from the users on compute nodes, and there are still some nodes like login nodes, data transfer nodes, and interactive debug queue nodes on which multiple simultaneous users are working.

Finally, we have also configured `pam_slurm` so that users can only ssh into compute nodes on which they have one or more jobs currently executing. This further restricts the activities a malicious actor can perform if they were to compromise a user's account.

*C. File Systems*

Most HPC systems have several file systems [22]: there are usually one or more central file systems that are mounted on all user-accessible nodes, which includes users' home directories and scratch storage. Also, every compute and login node usually has some user accessible local storage, which is only accessible locally on that node. Most computers implement a discretionary access control (DAC) model for access to resources, meaning the creator or owner of the file sets the access rules for that file [27]. We place limits on that discretion. Across all of this storage, our goal is that users are not be able to share data with any other user except through intentional use of an approved project group.

The intended way users can share code and data is only via approved project groups, which utilize UNIX group permissions. Each approved project group has one or more "data stewards", which are usually project leaders. These data stewards approve adding and deleting users in their groups, and they are responsible for the contents and the members' actions in their approved project groups. This assumes the standard user private group model is in use, which means every user's default group is a private group which contains only themselves. All user home directories are owned by root and group-owned by their user private group, preventing users from changing the permissions of their top level home directory. This also means that a user cannot give access to any file or directory in their home directory, because other users are not members of the user private group.

Along with these configurations and measures, we also wrote and deployed some Linux kernel patches to further restrict file system permissions [28]. These patches address concerns about the use of shared world-writable directories, such as the local `/tmp/` and `/dev/shm/` directories commonly found on Linux systems. It blocks the use of world bits for unprivileged users by setting a security mask (`smask`). This is similar to setting `umask 007`, but it is immutable and enforced (even on `chmod`). These patches also restrict the use of file access control lists to group members only, and a user cannot grant permission to a group unless they are a member of said group. We also wrote and upstreamed a Lustre patch to honor the `smask` setting[2] by replacing it's non-standard use of a direct read of the umask variable to the accessor function that is part of the standard kernel API and modified by our smask patch.

There are some situations that we have encountered where HPC support personnel who are not full administrators of the system need to set global file permissions. These include making widely-used datasets, AI models, and software tools (e.g., compilers, libraries, etc.) available to all users. Similar to `seepid`, we developed an `smask_relax` tool, so that the HPC support personnel can enter a new shell session with `smask 002` set so that they can set global read or execute file permissions on such data areas, and then leave the session [28].

While there has been other work on mandatory access control (MAC) model systems for use in multi-category or multi-level systems, the focus of this work has been on separating well-defined "projects" or "levels", and only been deployed in niche systems with these well-defined boundaries. These existing techniques have focused on "zoning" HPC resources into coarse buckets, often requiring network-level or node-level separation with few specific convergence points and not addressing fine-grained permissions within a bucket that a user has access to. They do not scale to thousands or tens of thousands of individual users and project groups. The limited deployment of these technologies has also negatively impacted their commercial viability and long-term supportability. An example of such technology is the Seagate ClusterStor Secure Data Appliance [29], which is no longer commercially available.

*D. Network*

Similar to the other areas, our goal for HPC networks is to only permit network connections between processes where the client and the server are running as the same user, with the ability to extend it to project groups on an opt-in basis. Firewalls are commonly used on servers and client computers, alike, to only make those network services available that are pre-approved by the applicable security policy. But rather than a traditional firewall based on the source and destination, along with defined ports, protocols, and services (PPS), we have developed and deployed a user-based firewall (UBF) for TCP and UDP traffic [30], [31]. We find this approach fits the normal use cases of HPC outlined in Section II much better than a traditional firewall. A traditional PPS firewall would have no way to make an intelligent decision about a traffic flow consisting of a novel application still in it's "version 0" phase of development, but this is no impediment to making user-based decisions.

Our UBF uses the IPTables NetFilter Queue module (`nfqueue`) to send new connection requests to a userspace daemon for decision. Only "new" connections are sent; IPTables connection tracking (`conntrack`) handles established connections. During the establishment of a new connection an ident [32]-like query is sent from the receiving system to initiating system to get user information, and the same query run locally. The connection is allowed if both the receiving and the initiating processes are owned by the same user or if the connector is a member of primary group of the listener process. The primary group of the listening process can be controlled via standard Linux tools such as `newgrp` or `sg`. The UBF also implicitly controls most IB/RDMA traffic because most such frameworks use TCP connection for setup. However, IB/RDMA traffic that uses the native IB connection manager for setup is not covered.

---
[2]Lustre issue LU-4746 was merged into in Lustre 2.7.0.

There has been other work at the application level on trying to secure the HPC network; most notable was an effort to encrypt all MPI traffic [33]. Greater detail of such efforts and their trade-offs are available in [30]. However, these efforts fall into the "Option 1" category we discussed in Section III and have the drawbacks discussed in that section.

There has been other work at the system level focused on bringing MAC to networking by labeling network traffic. IPTables has support to add security labels to traffic via `SECMARK` and `CONNSECMARK` based on traditional PPS firewall rules. Netflow [34] has the ability to pass tags based on IP packet option headers as described in RFC#1108 [35] and Commercial Internet Protocol Security Option (CIPSO) [36] that was never standardized, yet remains in use. Additionally, tags can be passed within an IPSec tunnel between hosts. Both of these options lack the flexibility to integrate with common HPC workflows. As discussed above, a PPS approach to network traffic is impractical on HPC systems, and as discussed in Section IV-C the coarse "level" controls of MAC-based approaches do not address the fine-grained access control within a bucket needed for HPC systems. Additionally, IPSec tunnels have significant overhead and negate many of the hardware offload network accelerations common in HPC systems.

### E. Web Portal/Gateway

LLSC systems enable application jobs that have web interfaces by forwarding the web connections from compute nodes to the user's laptop/desktop via an HPC portal [18]. This capability supports applications like Jupyter notebooks, Jupyter labs, TensorBoard, and more. It avoids ad-hoc port forwarding through SSH, SSL/TLS certificate warnings, and several classes of user misconfigurations. User authentication is required to connect to the HPC Portal and UBF connection rules (described in Section IV-D) are enforced, so that the entire connection path is authenticated and authorized. Providing this forwarding via our HPC portal allows us to launch applications with web interfaces on any compute node in any partition within the HPC system; they are not restricted to a small partition, which is commonly implemented at HPC centers.

The LLSC web portal/gateway was developed before other viable open source projects were available, but we continue to develop on it as a prototype for developing unique features in web portals/gateways and the integration with the other security features of our systems. These days, the most widely used web portal/gateway for HPC systems is Open OnDemand [37]. A good discussion of security challenges in web portals/gateways is included in this survey paper from 2019 [38].

### F. Accelerators

Accelerators, and specifically GPUs, do not use a traditional security model for data resident in memory. They have no concept of data ownership or data segmenting within the GPU [39], [40]. Therefore greater measures must be taken to associate GPUs to the jobs that have requested one or more GPUs through the scheduler. On LLSC systems, GPUs are assigned as a single-user resource. This is accomplished by modifying the permissions on relevant character special files in `/dev/` to allow only the user private group of the user allocated that GPU via the scheduler. With this method, GPUs that have not been assigned to a user are not visible at all. (This is not relevant when whole node scheduling with `pam_slurm` restrictions are in place.)

Also, GPUs do not clear their memory before reassignment to another job/user, because the GPU has no implicit way to know when it's being reassigned. Alas, the data of the previous user's job will remain in GPU memory and registers. We have implemented vendor-provided steps to clear the GPU, which are performed in the scheduler epilog script.

### G. Containers

Containers provide a convenient way to package applications, libraries for the application, and environment configurations within a share-able file. Containers come in two varieties: enterprise service containers and software encapsulation containers. There are some fundamental differences between enterprise service containers and software encapsulation containers (i.e., HPC containers). Enterprise service containers (e.g., Docker [41]) are lightweight versions of full-blown virtual machines. They were originally developed to run multiple services on a single server without all of the overhead of running multiple full virtual machines, and they were not intended to serve as software environment deployment mechanisms. Enterprise service containers encapsulate all of the application software and libraries to execute one or more services, much like full virtual machines, and they include network port virtualization, shared storage virtualization, USB peripheral virtualization, etc., which is common for multi-service deployments on shared servers. There is no need to carry around the entire operating system, libraries, etc., when running the container on a compatible/similar operating system; one just needs to address the differences with the underlying operating system and make sure none of the containers that are running in the same OS image are clashing with other network or storage name spaces. Enterprise service container environments assume that the container owner/executor has root privileges. On single-user laptops/desktops and enterprise service servers, it is quite common to have administrator privileges to manage these computers. However, HPC systems are inherently multi-user, and there are very strong requirements for administrators; general HPC users are forbidden to have any administrative privileges according to DoD security requirements.

The HPC community saw the utility of using containers as encapsulated software environment deployment mechanisms, particularly with the fragility of Python environments. HPC containers are heavyweight versions of Linux environment modules [42], [43]. There are several technologies in this category including Singularity/Apptainer [44], Charliecloud [45], and Shifter [46]. We compared these options, and we decided to support Singularity/Apptainer on LLSC systems because it had the best feature set and broadest adoption among HPC centers. To avoid granting any adminstrative privileges, users cannot create and populate their Singularity containers on the HPC system; they must use their own computer where they have some administrative privileges in order to do so.

Another distinguishing difference between enterprise service containers and HPC containers is that the unnecessary features are not implemented. HPC containers typically do not support USB peripheral virtualization, can only pass-through shared access to the host network stack, and often pass-through the host local and central file systems for their persistent storage. By disabling these features many security concerns of containers, including many of the configuration dependent concerns that would be different with each and every container, are eliminated. Additionally, all of the security features described in this paper pass through to the container as well.

Containers, however, do have some downsides. They open the HPC system up to other attack vectors including stale code and libraries and they are known to harbor vulnerable code [47]. This is not a product of the technology; rather it is a product of how containers are used. Secondly, because of the ease with which they can be shared among shared-group users, containers tend to get proliferated across central file systems by sharing, cloning, and modifying them. After a few years, there are just a lot of old, unused containers littering the home directories and shared group areas of central file systems. Users do not remember why they are still keeping them without a very disciplined container deployment and management process. We have found that shared installations of software applications are better managed by providing installed applications in shared group areas and enabling users to dynamically configure their environment to use the applications with Linux environment modules.

There are certainly situations for which Singularity containers are the best solution, and we do enable Singularity privileges to users and teams for which this is the case. For them, we make sure that they are aware of the advantages and disadvantages so that they can manage the software of their project more adroitly.

## V. RESULTS

By implementing these sets of security configurations and measures, opportunities for accidental data leakage between users are greatly reduced. There remain a few paths that still exist, including file names in world-writable directories (e.g., `/tmp/`, `/dev/shm/`), abstract namespace unix domain sockets, and direct IB verbs network communication. The configurations and measures described in this paper enhance reliability as well. Even if two users accidentally choose the same port number for a network service, they cannot crosstalk and corrupt each others data. On the whole, this limits the damage of misbehaving code and contains the extent of effect or "blast radius" of any issues to just that user's account. And for users, it looks like they're the only one on the HPC system.

Some users may find that the formalism of having HPC system staff create approved project groups for them and declaring the participants of the group to be bureaucratic and involves too much overhead. However, these are teaching opportunities for students, staff, and faculty alike to learn about restrictions, licenses, and requirements of handling data and software properly and legally. Many of them have not had to face these realities before.

The measures and technologies detailed in this paper have greatly improved the security, usability, and reliability of our systems for our users. By having these controls in place, and enforced at a system level, we have also been able to give the sponsors of the users' work much greater confidence that their data is secure and it won't be exposed to the entire userbase of the system due to a user misconfiguration or a mis-typed command.

## VI. CONCLUSION

Every HPC user is a software developer, but software development is not their primary domain of expertise, nor will it ever be. By enabling strong user separation at every point in the system, the confidentiality and integrity of the data is protected, and by reducing the burden on the user to worry about these things, the usability of the system is enhanced as well.

### ACKNOWLEDGEMENTS

The authors express their gratitude to Bob Bond, Alan Edelman, Peter Fisher, Jeff Gottschalk, Chris Hill, Charles Leiserson, Kirsten Malvey, Heidi Perry, Steve Rejto, Mark Sherman, and Marc Zissman for their support of this work.

## APPENDIX
### REPRODUCIBILITY APPENDIX

While many of the security measures mentioned in this paper are configuration settings, technology choices, and processes, there are two code bases that are mentioned in the text and are available on `github.com`.

The first code repository is the File Permission Handler [28], which was mentioned in Section IV-C. The code in this repository consists of two Linux kernel patches and a PAM module that implement restrictions on the file permissions available to users on an HPC system. When used with an smask setting of `007` on a system which implements the user private group scheme, these changes effectively prevent users sharing data via the filesystem unless they are both members of the same supplemental group.

The second code repository is the User-Based Firewall (UBF) [31], which was mentioned in Section IV-D. The code in this repository consists of two daemons and two Apache httpd plug-ins that, together with specific Linux nftables firewall rules, make up the UBF. The UBF can act on both tcp and udp connections, and would normally be configured (via iptables or other firewall utilities) to inspect connections on ports numbered 1024 and above. The ruleset implemented only permits a connection when the connecting and listening processes are running as the same user, or the connecting process is a member of the primary group (`egid`) of the listening process. When used on a system which implements the user private group scheme and disables protocols other than TCP and UDP, these changes effectively prevent users sharing data via the network unless they are both members of the same supplemental group.

While the UBF does not directly affect code using Infiniband verbs or remote direct memory access (RDMA), many such applications use a TCP connection as a control channel to set up their Infiniband queue pairs (QPs) and thus can be effectively controlled by the UBF. This does not prevent applications from using the connection manager (CM) directly to set up their QPs, and any application that does this would not be controlled by the UBF.